\documentclass[a4paper,usenames,dvipsnames,11pt]{article}
\pdfoutput=1

\usepackage{jheppub}
\usepackage{slashed}
\usepackage{mathrsfs,booktabs,multirow,tabularx}
\usepackage{stmaryrd}
\usepackage{xspace}
\usepackage{fancyvrb}
\usepackage[makeroom]{cancel}
\usepackage{amsmath}    
\usepackage{amssymb}    %
\usepackage{graphicx}   
\usepackage{verbatim}   
\usepackage{lscape}
\usepackage{subfig}
\usepackage{listings}

\def\beq{\begin{equation}}
\def\beqn{\begin{eqnarray}}
\def\eeq{\end{equation}}
\def\eeqn{\end{eqnarray}}
\def\abs#1{\left|#1\right|}

\def\PDF#1#2{\Gamma_{\!#1/#2}}
\def\ePDF#1{\Gamma_{\!#1}}

\newcommand\sss{\scriptscriptstyle}
\newcommand\mydot{\!\cdot\!}

\newcommand\aW{\alpha_{\sss W}}
\newcommand\as{\alpha_{\sss S}}
\newcommand\aem{\alpha}

\newcommand{\bq}{\bar{q}}

\newcommand{\epem}{e^+e^-}
\newcommand{\mpmm}{\mu^+\mu^-}
\newcommand{\lp}{e^+}
\newcommand{\lm}{e^-}

\newcommand{\mup}{\mu^+}
\newcommand{\mum}{\mu^-}
\newcommand{\mupm}{\mu^{\pm}}

\newcommand{\ord}{{\cal O}}

\newcommand\QCD{{\rm QCD}}
\newcommand\QED{{\rm QED}}

\newcommand\NF{N_{\sss F}}
\newcommand\CF{C_{\sss F}}

\newcommand\CA{C_{\sss A}}

\newcommand\Nl{N_l}
\newcommand\Nu{N_u}
\newcommand\Nd{N_d}
\newcommand\bzQED{\beta_0}
\newcommand\boQED{\beta_1}
\newcommand\hsig{\hat{\sigma}}

\newcommand\APmat{{\mathbb P}}

\newcommand\Eop{{\mathbb E}}
\newcommand\Mmat{{\mathbb M}}

\newcommand\MSb{\overline{\rm MS}}

\newcommand{\pt}{p_{\sss T}}

\newcommand\mZ{m_{\sss Z}}

\newcommand{\LQCD}{\Lambda_{\rm QCD}}

\title{The muon parton distribution functions}

\author[a]{S. Frixione,}
\affiliation[a]{INFN, Sezione di Genova, Via Dodecaneso 33, I-16146, 
Genoa, Italy}

\author[b]{G. Stagnitto}
\affiliation[b]{Physik-Institut, Universit\"at Z\"urich, 
Winterthurerstrasse 190, CH-8057 Z\"urich, Switzerland}

\emailAdd{Stefano.Frixione@cern.ch}
\emailAdd{giovanni.stagnitto@physik.uzh.ch}

\abstract{
We compute the Parton Distribution Functions (PDFs) of the unpolarised
muon for the leptons, the photon, the light quarks, and the gluon.
We discuss in detail the issues stemming from the necessity of evaluating 
the strong coupling constant at scales of the order of the typical hadron 
mass, and compare our novel approach with those currently available in
the literature. While we restrict our phenomenological results to be
leading-logarithmic accurate, we set up our formalism in a way that
renders it straightforward to achieve next-to-leading logarithmic
accuracy in the QED, QCD, and mixed QED$\times$QCD contributions.
}


\preprint{
\begin{flushright}
ZU-TH 49/23\\
\today
\end{flushright}
}

\begin{document}
\maketitle
\flushbottom

\section{Introduction\label{sec:intro}}
The post-LHC generation of high-energy physics experiments will
most likely be at lepton colliders. The vast majority of
extensive feasibility studies for such colliders have been, and are,
carried out by assuming electron/positron beams, with either a circular
or a linear geometry. An interesting alternative option is that of 
a $\mpmm$ collider; while the technical difficulties of controlling
and accelerating muon beams remain significant, so are the benefits
stemming from building such a machine. Among these, a prominent one
is the possibility of achieving much larger c.m.~energies w.r.t.~those
relevant to $\epem$ colliders with much smaller rings; to be definite, 
we shall assume that at the typical $\mpmm$ accelerator one will have 
\mbox{$\sqrt{s}=\ord(1-10~{\rm TeV})$}.

Because of the large collider energy, the incoming muons are expected
to copiously radiate before initiating the hard process; in fact, the
particles produced during this radiation phase may eventually undergo
hard collisions themselves. Thus, the situation is quite analogous to
that of hadronic collisions, and indeed production processes can be described
in the same manner, namely by means of a collinear-factorisation formula
\beq
\sigma_{\mpmm\to X}(s)=\sum_{ij}\int dz_1dz_2
\PDF{i}{\mup}(z_1)\PDF{j}{\mum}(z_2)
\hsig_{ij\to X}(z_1z_2s)\,,
\label{fact0}
\eeq
in which process-independent Parton Distribution Functions (PDFs,
$\PDF{i}{\mup}$ and $\PDF{j}{\mum}$) are convoluted with process-specific
short-distance cross sections ($\hsig_{ij\to X}$), in which partons
($i$ and $j$) that emerge from the muon beams collide in the hard process. 
This picture is of course valid for $\epem$ collisions as well; the difference 
between the two is that the larger $\mpmm$ c.m.~energies imply that the 
muon PDFs {\em can} be probed at values of their arguments (the 
``Bjorken $x$'', $z_1$ and $z_2$ in eq.~(\ref{fact0})) which are much 
smaller than their electron counterparts, since when the outgoing system $X$ 
is produced in the central region of a detector with an invariant mass 
squared equal to $Q^2$ , then \mbox{$z_{1,2}\sim\sqrt{Q^2/s}$}. With that being
said, the peculiarity of lepton colliders w.r.t.~to hadron ones is
that the majority of hard collisions still result in large-invariant mass
systems, and are due to partons whose identities are the same as those 
of the respective beams, i.e.~$i=\mup$ and $j=\mum$,
since \mbox{$\PDF{\mupm}{\mupm}(z)\sim\delta(1-z)$}. Conversely, when
$Q^2\ll s$, the production process is generally initiated by partons
whose identities are different w.r.t.~those of the respective beams
($i\ne\mup$ and $j\ne\mum$), since for these partons the PDFs peak at 
small $z$'s. We point out that the set of such partons includes the 
strongly-interacting ones, i.e.~the quarks and the gluon.

It is therefore important, in the context of $\mpmm$-collider studies,
that muon PDFs be available for all possible parton types. This necessity
has recently been addressed by two different groups~\cite{Han:2020uid,
Han:2021kes,Garosi:2023bvq}, which have essentially used the same
approach. In this paper we follow a strategy that differs from that
of refs.~\cite{Han:2020uid,Han:2021kes,Garosi:2023bvq} in three major 
aspects; we shall comment in detail about these in sect.~\ref{sec:mot}.
The technical features of our work are discussed in sect.~\ref{sec:tech},
and the resulting PDFs are presented in sect.~\ref{sec:res}, where we
also give sample predictions relevant to dijet production. Finally,
we conclude and give an outlook on future work in sect.~\ref{sec:concl}.

\section{Motivations\label{sec:mot}}
There are three defining features of the PDFs of refs.~\cite{Han:2020uid,
Han:2021kes,Garosi:2023bvq}. Firstly, the strong-coupling constant is
set equal to zero for all scales below a certain threshold, 
\mbox{$\as(\mu)=0$} for \mbox{$\mu\le Q_0$}, while above that threshold
a perturbative RGE-evolved form is adopted\footnote{$Q_0$ is denoted
by $\mu_{\rm QCD}$ and $Q_{\rm QCD}$ in ref.~\cite{Han:2021kes} and
ref.~\cite{Garosi:2023bvq}, respectively.}. Secondly, the $W$ and the 
$Z$ bosons are regarded as partons, and included in the evolution above the
EW scale. Thirdly, the evolution starts from leading-order (LO) initial
conditions, and is leading-logarithmic (LL) accurate. We shall now 
comment on these aspects, w.r.t.~which our work has significant differences.

As far as the treatment of strong interactions is concerned, one must
bear in mind that lepton PDFs are naturally evolved starting from
a scale $\mu_0$ of the order of the lepton mass, and ideally strictly
equal to the mass itself (lest spurious logarithms are generated in
the evolution~\cite{Bertone:2019hks}). By considering strongly-interacting
partons, one is therefore in principle forced to evaluate $\as$ at scales
of the same order as $\Lambda_{\rm QCD}$, thus outside of the perturbative
domain. As was said before, this problem is circumvented in 
refs.~\cite{Han:2020uid,Han:2021kes,Garosi:2023bvq} by simply switching
QCD off below a scale \mbox{$Q_0=\ord(1~{\rm GeV})$}. Obviously, this is
a purely technical procedure: no dynamical mechanism is invoked that could
result in this behaviour. More concerning, a strategy of this kind can be 
seen as equivalent to a renormalon-inspired approach whereby $Q_0$ is a free 
parameter whose value is completely unconstrained, and fixed a posteriori 
in an observable-dependent manner (see e.g.~ref.~\cite{Webber:1998um} 
for a discussion that focuses on cross sections). In the context of PDF
evolution it is unclear whether such a potential dependence on the observable
destroys PDF universality and, more pragmatically, how the value of $Q_0$
should be chosen (ref.~\cite{Han:2021kes} sets $Q_0=0.5$~GeV and
considers $Q_0=0.7$~GeV as an alternative option, whereas in 
ref.~\cite{Garosi:2023bvq} variations in the range 
\mbox{$0.52\le Q_0\le 1$}~GeV are explored, the default
value being $Q_0=0.7$~GeV). In this paper, we instead adopt a simple 
analytical parametrisation~\cite{Webber:1998um} of $\as$ which is valid 
for any non-null scale value and is inspired by a dispersive approach; we
shall discuss this in sect.~\ref{sec:QCD}.

For what concerns the heavy EW vector bosons $V=W,Z$, we remind the reader 
that the argument for treating them as partons amounts essentially to saying
that, since $m_V^2\ll s$, one is in a regime of collinear dominance, where 
the boson mass is negligible, and its role is thus similar to that of any
light particle. In fact, the actual condition is whether $m_V^2\ll Q^2$
or not, where $Q^2$ is the invariant mass squared of the system produced 
in the hard collision. However, as was discussed in sect.~\ref{sec:intro},
for partons different from the muon (therefore including the vector bosons),
the cross section associated with invariant masses of the order of the
c.m.~energy is very small, being strongly suppressed by the $z\to 1$ 
behaviour of the PDFs themselves. In order to be slightly more quantitative,
let us consider the typical figure of merit of collinear physics, which
in the case of a vector boson $V$ reads \mbox{$\aW/\pi\log(Q^2/m_V^2)$}.
For this to have, say, the same value ($0.067$) as its analogue 
\mbox{$\aem/\pi\log(s/m_e^2)$} relevant to an electron in a
$\sqrt{s}=350$~GeV collision (i.e.~at the FCC-ee) one needs 
$\sqrt{Q^2}\sim 2$~TeV.
The c.m.~energy required for this value of $Q^2$ to correspond to a 
non-negligible cross section depends of course on the process considered,
but an average rough estimate would lead to tens of TeV's (see 
e.g.~fig.~1 of ref.~\cite{Han:2021kes}). This is in fact compatible
with observations made elsewhere (see e.g.~refs.~\cite{Bauer:2017isx,
Ruiz:2021tdt}) that for true collinear dominance to be achieved in
the case of massive vector bosons, and thus for resummation to be
necessary, the energies involved need to be much larger than those
suggested by naive logarithmic counting. We also point out that 
with $\sqrt{Q^2}=2$~TeV one has \mbox{$m_V^2/Q^2\sim 2\mydot 10^{-3}$};
for comparison, \mbox{$(\aW/\pi)^2\sim 1.1\mydot 10^{-4}$}: in other 
words, power-suppressed effects are larger than the coupling-constant
factor of a \mbox{$2\to 2+n$} process relative to a \mbox{$2\to n$} 
one\footnote{Roughly speaking, power-suppressed effects are neglected
in a PDF-based approach, but exactly accounted for in matrix-element
computations. A \mbox{$2\to 2+n$} process is representative of a
VBF-type reaction, that may be described by employing two EW-boson 
PDFs, whose final state is produced at the leading order by a 
\mbox{$2\to n$} reaction.}. Finally, we mention the fact that the
inclusion of EW heavy bosons in the set of partons implies that
the evolution above the EW is carried out in the unbroken EW phase.
The practical consequence of this is that there will exist PDFs associated 
with non-standard objects, namely the ``$Z/\gamma$ interference'',
that are meant to be convoluted with interference matrix elements
(i.e.~whose amplitudes on the left- and right-hand side of the cut
differ); matrix elements of the latter kind are presently not readily 
available in codes that compute short-distance cross sections
(see however refs.~\cite{Bothmann:2020sxm,Pagani:2021vyk,Pagani:2023wgc}
for recent progress).
For all these reasons, in our approach we do not treat the EW heavy
vector bosons as partons, and we do not include them in the evolution
of our muon PDFs, regardless of the scale considered.

Finally, refs.~\cite{Han:2020uid,Han:2021kes,Garosi:2023bvq} work at the 
LO+LL precision. This implies that the only non-null initial condition is 
that of the muon (equal to $\delta(1-z)$), with all of the other PDFs
generated dynamically, with evolution equations based on $\ord(\aem)$
and $\ord(\as)$ splitting kernels. While presently this is a sufficiently
pragmatic choice (to which we shall also adhere for our phenomenological
predictions), a future increase of accuracy may prove very demanding,
since it will entail the computation of the next-to-leading order (NLO)
splitting kernels relevant to the heavy-boson sector. Conversely,
our approach is based on a simpler QED+QCD evolution, for which 
all of the required kernels and initial conditions are available 
at least to NLO; we can thus easily obtain NLO+NLL-accurate predictions.
Importantly, an additional benefit of that is the capability of studying 
explicitly the small-$x$ behaviour of the PDFs.

We conclude this section with an aside that concerns the assumptions
on small-scale physics that enter the muon-PDFs determination. If one
focuses on the strongly-interacting partons, it may be tempting to
describe them as the convolution of the {\em photon} quark and gluon
PDFs with a function that gives the collinear-photon content of
the muon. The latter would be the Weizs\"acker-Williams
function~\cite{vonWeizsacker:1934nji,Williams:1934ad}, or any other
Equivalent Photon Approximation (EPA), while the former ones would
be taken from available fits (for those currently relevant, see
e.g.~refs.~\cite{Gluck:1991jc,Schuler:1995fk,Schuler:1996fc,Cornet:2004nb,
Slominski:2005bw}). The underlying idea here is that, thanks to their
data-driven determinations, photon PDFs would bypass the necessity
of a theoretical modeling of the small-scale behaviour of $\as$.
Unfortunately, such a strategy suffers from severe drawbacks that
prevent one from using it in practice. In particular, the quality
of photon-PDF fits is generally rather poor, thus inducing systematic
uncertainties comparable to or larger than those stemming from the
assumptions on $\as$ in the infrared region. Furthermore, EPAs cannot describe
the full complexity of the photon content of a lepton at arbitrary scales
(for a discussion specific to this point, see e.g.~sect.~4.2.3 of
ref.~\cite{Frixione:2019lga}). In addition to this, it is necessary to
bear in mind that the partonic contents of the muon are still predominantly
given by the muon itself and the photon, while the presence of the
other charged leptons must be taken into account too. It is unclear
how to combine the evolution of these partons with that
of the strongly-interacting partons originating from the mechanism
sketched above. Even if successful, such a combination would entail
evaluating the photon PDFs at scales of the order of the muon mass, which
poses the very problem one sought to avoid in the first place when
introducing these PDFs; obviously, this is true regardless of the
quality of the photon-PDF fits, and thus applies even if that quality
were to improve in the future.
For all of these reasons, we deem that the usage of photon PDFs is not
adequate for obtaining reliable muon PDFs, and we shall not discuss
this approach any further in this paper.

\section{Technicalities\label{sec:tech}}
We obtain the muon PDFs by following the same procedure as in
refs.~\cite{Bertone:2019hks,Bertone:2022ktl}: the evolution equations
are written in terms of an evolution operator and of the initial conditions
in Mellin space, solved there, and inverted numerically back to
the $z$ space. The only differences w.r.t.~what has been done in 
refs.~\cite{Bertone:2019hks,Bertone:2022ktl} are: {\em a)}~the inclusion
of the $\ord(\as)$ splitting kernels; {\em b)}~the addition of the gluon 
as an active parton. Furthermore, we neglect the contributions of
the $\ord(\aem^2)$ splitting kernels, since these are on the same
footing as their $\ord(\as^2)$ and $\ord(\aem\as)$ counterparts, which
we ignore here. These changes do not entail any conceptual modifications
w.r.t.~refs.~\cite{Bertone:2019hks,Bertone:2022ktl}, and only a few minor 
practical ones; more details are given in sect.~\ref{sec:evol}.
For consistency, we use LO initial conditions, but we note that the 
$\ord(\aem)$ (i.e.~NLO) ones computed in ref.~\cite{Frixione:2019lga}
do not require any changes when strong interactions are included 
(the effects of the latter being of $\ord(\aem\as)$ or higher). 
We take into account all parton-mass thresholds: each parton 
participates in the evolution only at and above the corresponding 
threshold, set equal to its mass (whose value is in turn that of the 
PDG~\cite{Zyla:2020zbs}), as is done in ref.~\cite{Bertone:2022ktl};
the starting scale of the evolution $\mu_0$ is set equal to the muon mass.
Finally, lest the stability of the results be degraded, at large $z$'s we 
switch from the numerical to the analytical solution, and are thus able to
obtain a smooth behaviour in the whole $z\in (0,1)$ range; we point
out that the analytical $z\to 1$ solutions of refs.~\cite{Bertone:2019hks,
Bertone:2022ktl} are unchanged when including strong interactions
(both at the LL and the NLL accuracy).

\subsection{QCD at small scales\label{sec:QCD}}
The setup just described, in particular the fact that $\mu_0=m_\mu$
and the inclusion of the $\ord(\as)$ splitting kernels, requires
that $\as(\mu)$ be computed for any $\mu\ge m_\mu$. As was anticipated
in sect.~\ref{sec:mot}, we are able to do so by adopting the analytical
parametrisation of $\as$ of ref.~\cite{Webber:1998um}. We point out that, 
as for all approaches that assume a universal small-scale behaviour of 
$\as$, this is expected to induce power-suppressed corrections of
\mbox{$\ord(\LQCD^{2p}/Q^{2p})$} to observables which, given the
typical values of $Q^2$ relevant to $\mpmm$ collisions, we shall 
safely neglect here. At the same time, using e.g.~a dispersive 
approach~\cite{Dokshitzer:1995qm}, one sees that the singular behaviour
of QCD structure functions is unaffected~\cite{Dasgupta:1996hh} by the
small-scale behaviour of $\as$, and thus that the evolution equations 
are unchanged.

As far as the strong coupling constant is concerned, for a better
phenomenological treatment we use the two-loop QCD $\beta$ 
function\footnote{In principle, at this order the evolutions of $\as$ 
and $\aem$ mix; we ignore such mixing effects here, which are in any case 
expected to be negligible in practice (see 
e.g.~ref.~\cite{xFitterDevelopersTeam:2017fxf}).\label{ft:asaem}}:
\beq
\beta_{\QCD}(\as)=-b_0\as^2-b_1\as^3+\ord(\as^4)\,,
\label{beta2l}
\eeq
with
\beqn
b_0&=&\frac{11\CA-2\NF}{12\pi}\,,
\\
b_1&=&\frac{17\CA^2-5\CA\NF-3\CF\NF}{24\pi^2}\,,
\eeqn
and the number of flavours $\NF$ is understood to be that relevant to
the scale range one is working in. Thus, $\NF=5$ for $\mu>m_b$,
$\NF=4$ for \mbox{$m_c<\mu<m_b$}, and $\NF=3$ for \mbox{$\mu_0<\mu<m_c$}
(note that following refs.~\cite{Bertone:2022ktl,Zyla:2020zbs} we assume
$m_s<m_\mu$). The exact solution for $\as(\mu)$ of the RGE stemming from
eq.~(\ref{beta2l}) is:
\beq
\int_{\as^f}^{\as^{\rm RGE}(\mu)}\frac{da}{\beta_{\QCD}(a)}=
\log\frac{\mu^2}{\mu_f^2}\,,
\label{asRGE}
\eeq
where $\as^f$ is the input value of the coupling constant at the reference
scale $\mu_f$. The coupling constant to be used in the full scale
range \mbox{$\mu_0<\mu<\infty$} is defined as the sum of a perturbative
and a non-perturbative part:
\beq
\as(\mu)=\as^{\rm pert}(\mu)+\as^{\rm np}(\mu)\,.
\label{asdef}
\eeq
As far as the former is concerned, we use:
\beq
\as^{\rm pert}(\mu)=\left\{
\begin{array}{ll}
1\Big/\Big(b_0\log\big[y(\mu)\big]\Big)
   &\phantom{aaaaaa}\mu<\mu_{\rm pert} \\
\as^{\rm RGE}(\mu)                &\phantom{aaaaaa}\mu>\mu_{\rm pert} \\
\end{array}
\right.\,,
\label{asp1}
\eeq
while for the non-perturbative part, we use:
\beqn
\as^{\rm np}(\mu)&=&\left\{
\begin{array}{ll}
\frac{1}{b_0}a^{\rm np}(y(\mu)) &\phantom{aaaaaa}\mu<\mu_{\rm pert} \\
0                               &\phantom{aaaaaa}\mu>\mu_{\rm pert} \\
\end{array}
\right.\,,
\label{asnp1}
\\
a^{\rm np}(y)&=&\frac{y+b}{(1-y)(1+b)}
\left(\frac{1+c}{y+c}\right)^p\,,
\label{asnp2}
\eeqn
where $b$, $c$, and $p$ are free, low-energy parameters, The
auxiliary functions $x(\mu)$ and $y(\mu)$ are:
\beq
y(\mu)=x(\mu)\left[1+
\left(\frac{\log x(\mu)}{2\pi}\right)^2\right]^{\frac{b_1}{2b_0^2}}\,,
\;\;\;\;\;\;\;\;
x(\mu)=\frac{\mu^2}{\LQCD^2}\,,
\label{xyfun}
\eeq
where $\LQCD$ is implicitly defined as follows:
\beq
\as^f=\frac{1}{b_0\log\big[y(\mu_f)\big]}+
\frac{1}{b_0}\,a^{\rm np}(y(\mu_f))\,.
\label{LQCDdef}
\eeq
All of the above can be 
restricted to one-loop accuracy by formally setting $b_1=0$ everywhere.

We point out that the upper forms in eqs.~(\ref{asp1}) and~(\ref{asnp1})
guarantee a continuous and smooth behaviour of $\as(\mu)$ at 
the Landau pole $\mu=\LQCD$, so they must be adopted at small scales. 
Conversely, at large scales the difference w.r.t.~the exact RGE solution 
may be noticeable, and for this reason we have introduced the scale
$\mu_{\rm pert}$ at which we switch from one functional form
to the other. In practice, it is convenient to set $\mu_{\rm pert}$
equal to one of the quark-mass thresholds: of the two possible sensible
options ($m_c$ and $m_b$) we choose the former: $\mu_{\rm pert}=m_c$
henceforth. With this, we proceed as follows: we start by setting 
$\as^f=0.118$ at \mbox{$\mu_f=\mZ\equiv 91.18$}~GeV; with $\NF=5$, this 
determines $\as^{\rm RGE}(\mu)\equiv\as(\mu)$ for any $\mu\ge m_b$ by
means of eq.~(\ref{asRGE}). When then iterate the procedure 
by setting $\as^f=\as^{\rm RGE}(\mu_f)$ at $\mu_f=m_b$, so as to obtain 
$\as(\mu)$ in \mbox{$m_c\le\mu<m_b$} by solving again eq.~(\ref{asRGE})
with $\NF=4$. We now switch to the upper forms in eqs.~(\ref{asp1})
and~(\ref{asnp1}) and, by setting $\mu_f=m_c$, $\as^f=\as^{\rm RGE}(\mu_f)$,
and $\NF=3$, we determine $\LQCD$ with eq.~(\ref{LQCDdef})\footnote{At
$\mu_f=m_c$, the second term on the r.h.s.~of eq.~(\ref{LQCDdef}) has
a relative impact on the determination of $\LQCD$ of $\ord(10^{-4})$,
and can be thus ignored for all practical purposes. This is not true 
any longer when matching conditions are imposed at the lower strange-
and down-quark thresholds.}, after fixing the low-energy parameters as 
is indicated in the ``def'' column of 
table~\ref{tab:bcppars}~\cite{Webber:1998um}. This allows us to obtain
$\as(\mu)$ in \mbox{$m_s\le\mu<m_c$}, and therefore to iterate the
procedure just described, with the $\NF=2$ value of $\LQCD$ determined
after setting $\mu_f=m_s$ and $\as^f=\as(\mu_f)$. The procedure terminates
with the matching conditions imposed at the down-quark threshold. In other 
words, we do not impose any matching conditions at the up-quark threshold 
(the up quark being the lightest of the quarks with our settings), and 
therefore the parameters of $\as$ have the same values for 
\mbox{$\mu\in (0,m_u)$} as for \mbox{$\mu\in (m_u,m_d)$}.
We point out that the computation of $\as(\mu)$ for any value $\mu<m_\mu$ 
would be (indirectly) relevant to the muon PDFs only if we aimed 
at extracting the low-energy parameters of eq.~(\ref{asnp2}) from
data (see eq.~(\ref{Fabdef})), which we do not in this work.
On the other hand, what was done above allows one to use, without
any modifications, this form of  $\as(\mu)$ when including QCD effects 
in the PDFs of the electron (which we shall not do here).
\begin{table}[th!]
  \begin{center}    
    \begin{tabular}{|c|ccccc|}
      \hline
  & def & do1 & do2 & up1 & up2 \\
      \hline
$b$ & 0.25 & 0.2125 & 0.2875 & 0.2875 & 0.2125\\
$c$ & 4    & 4.6    & 4.6    & 3.4    & 3.4\\
$p$ & 4    & 4      & 4      & 4      & 4\\
     \hline
    \end{tabular}
  \end{center}
  \caption{
\label{tab:bcppars} Values of the non-perturbative parameters
that appear in eq.~(\ref{asnp2}). Those in the first column are
the default ones; the other columns report the variations we consider
in this work.}
\end{table}

There are a number of practical, as opposed to conceptual, differences 
w.r.t.~what has been done in ref.~\cite{Webber:1998um}. In particular,
we work here at the two-loop level, rather than at the one-loop one;
we implement mass thresholds and continuity conditions at them; and
we use the exact RGE solution for the perturbative part of coupling 
constant for scales larger than the charm mass. Conversely, the 
low-energy parameters $b$, $c$, and $p$ have the same default values 
as in ref.~\cite{Webber:1998um}. This does not need to be so, and in
principle these parameters should be fitted to experimental data.
This is outside of the scope of the present work; however, we
remark that in ref.~\cite{Webber:1998um} a general compatibility
(within $\ord(20\%)$ uncertainties) with event-shape and structure 
functions data is established by computing the $(a,b)=(1,1)$ and $(2,1)$ 
moments of $\as$ according to the definition
\beq
F_{a,b}(\mu)=\frac{a}{\mu^a}\int_0^\mu\frac{d\kappa}{\kappa}\kappa^a
\Big[\as(\kappa)\Big]^b
\label{Fabdef}
\eeq
at $\mu=2$~GeV. In view of this, we have computed the same moments
for different scenarios; we report the results in table~\ref{tab:asmom}.
\begin{table}[th!]
  \begin{center}    
    \begin{tabular}{|c|cccccccc|}
      \hline
  & LO(a) & LO(b) & LO(c) & def & do1 & do2 & up1 & up2 \\
      \hline
$F_{1,1}(2~{\rm GeV})$ & 0.511 & 0.417 & 0.543 & 
                         0.469 & 0.440 & 0.455 & 0.504 & 0.487 \\
$F_{2,1}(2~{\rm GeV})$ & 0.450 & 0.344 & 0.497 & 
                         0.400 & 0.393 & 0.395 & 0.406 & 0.404 \\
     \hline
    \end{tabular}
  \end{center}
  \caption{
\label{tab:asmom} Values of the first two moments of $\as$ for 
various settings of the low-energy parameters. See the text
for details.} 
\end{table}
The LO(a) column results from performing the same computation as in
ref.~\cite{Webber:1998um}: one-loop $\beta$ function, with $\NF=3$ evolution 
from $\as(\mZ)=0.118$; we find perfect agreement with table~1 of that paper. 
For the LO(b) and LO(c) columns we also use a one-loop $\beta$ function,
but we now implement continuity conditions at the charm and bottom
thresholds, starting with an $\NF=5$ evolution at the $Z$ mass and
setting there $\as(\mZ)=0.118$ and $\as(\mZ)=0.135$, respectively.
Since $\as(\mZ)=0.118$ is also the initial condition chosen in
ref.~\cite{Webber:1998um}, the difference between the LO(a) and
LO(b) results is entirely due to using threshold conditions and
the corresponding number of flavours in the latter case, which gives
significantly smaller values for the moments. Column LO(c) shows that,
in order to increase these moments with a variable-flavour-number evolution,
one needs to start from a much larger $\as(\mZ)$ value; this is actually
consistent with what is observed when fitting one-loop-evolved hadronic 
PDFs (see e.g.~ref.~\cite{Yan:2022pzl}). 
The remaining columns in table~\ref{tab:asmom} report the results
obtained with the two-loop, threshold-matched evolution discussed at
the beginning of this section; we see that those obtained with the
default parameters are slightly smaller than, but within 10\% of, 
those of ref.~\cite{Webber:1998um}.

The bottom line is that, for the present phenomenological needs, the
default values of the low-energy parameters that enter the non-perturbative
part of $\as(\mu)$ can be set as is done in ref.~\cite{Webber:1998um}.
Again, the ranges in which they can vary are best established by means
of comparisons to data. Absent those, a theoretical argument is that
of choosing them in a way that prevents $\as(\mu)$ from being larger
than one or smaller than zero (neither of which conditions is guaranteed
a priori by eq.~(\ref{asnp2})). By using this argument while fixing
the value of $p$ (so as to have an extremely sharp decrease of 
power-suppressed effects) we end up by tolerating $\pm 15\%$ variations
w.r.t.~the default values of the $b$ and $c$ parameters. These variations
can be combined in the same or in opposite directions, resulting in
the four possible scenarios denoted by ``do1'', ``do2'', ``up1'', and
``up2'' in table~\ref{tab:bcppars}. These lead to the predictions for
the first two moments of $\as$ reported in the four rightmost columns
of table~\ref{tab:asmom}.
\begin{figure}[thb]
  \begin{center}
  \includegraphics[width=0.95\textwidth]{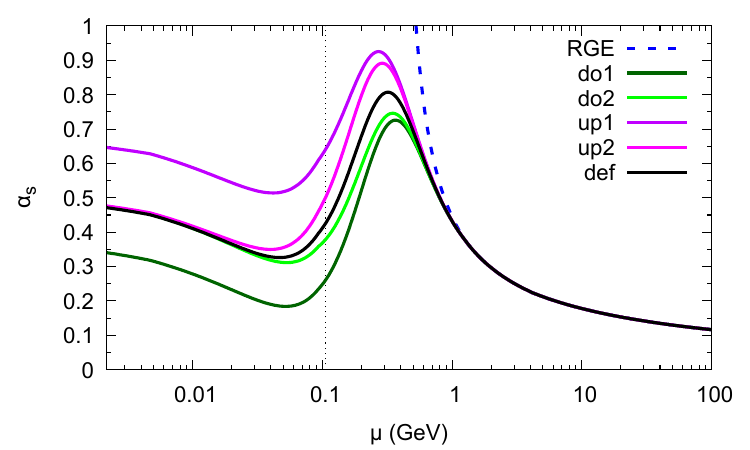}
\caption{\label{fig:alphaS} Strong coupling constant as a function
of the scale (in GeV), for various choices of the low-energy parameters.
The vertical dotted line indicates the value of the muon mass: for this
work, only the values to its right are relevant.
See the text for details.
}
  \end{center}
\end{figure}

The behaviour of $\as(\mu)$ as a function of $\mu$ is shown in
fig.~\ref{fig:alphaS}, for $\mu>m_u$\footnote{We stress again that 
$\as(\mu)$ is a regular function for any $\mu\ge 0$. For example,
in the case of the solid black line we obtain $\as(0)\simeq 0.594$.}. 
The five solid lines are the results for 
$\as(\mu)$ as is defined in eq.~(\ref{asdef}), while the dashed
line is the result of the solution $\as^{\rm RGE}(\mu)$ of the 
perturbative RGE equation~(\ref{asRGE}). The solid lines correspond
to the different choices of the non-perturbative parameters of
table~\ref{tab:bcppars}: black for default, dark green for do1, light green
for do2, dark magenta for up1, and light magenta for up2. From the figure and
the results of table~\ref{tab:asmom} we infer that the low-energy parameter
variations we consider in this paper are quite conservative, and likely
overestimate the true uncertainty of the analytical model adopted
for the strong coupling constant.

\subsection{PDF evolution\label{sec:evol}}
Following the notation introduced in ref.~\cite{Bertone:2022ktl}, 
we adopt an evolution basis with a 5-dimensional singlet sector, 
plus 13 non-singlet functions.
For factorisation and renormalisation schemes both chosen equal to $\MSb$,
we write the splitting kernels relevant to the non-singlet and singlet
evolution as follows
\beqn
P(x,\mu)=\sum_{i=0}^\infty\sum_{j=0}^\infty
\left(\frac{\as(\mu)}{2\pi}\right)^i
\left(\frac{\aem(\mu)}{2\pi}\right)^j
P^{(i,j)}(x)\,,
\label{APscex}
\\
\APmat(x,\mu)=\sum_{i=0}^\infty\sum_{j=0}^\infty
\left(\frac{\as(\mu)}{2\pi}\right)^i
\left(\frac{\aem(\mu)}{2\pi}\right)^j
\APmat^{(i,j)}(x)\,,
\label{APmatex}
\eeqn
respectively. As is customary, the coefficients on the r.h.s.~of
eqs.~(\ref{APscex}) and~(\ref{APmatex}) can be further decomposed into
their valence and sea components, by means of which one defines in turn
the standard $\pm$ combinations.

With this notation, each of the non-singlet PDFs evolve according to:
\beq
\frac{\partial\ePDF{a}}{\partial\log\mu^2}=
P_a\otimes \ePDF{a}
\label{evolNS}
\eeq
which applies to the following $13$ cases when $\mu>m_b$ (since there 
we consider the evolution with $\Nl=3$ lepton, $\Nu=2$ up-type quarks, 
and $\Nd=3$ down-type quarks, denoted collectively by $f$ in the last
line of eq.~(\ref{tableNS})), 
\beq
\begin{array}{rclcc}
\ePDF{a} & & 
& \phantom{aaa}\longleftrightarrow \phantom{aaa} &
P_{a}
\\
\hline
\ePDF{l2}&\;=\;&\ePDF{\lm}+\ePDF{\lp}-\left(\ePDF{\mu^-}+\ePDF{\mu^+}\right)
& \phantom{aaa}\longleftrightarrow \phantom{aaa} &
P_{l}^+
\\
\ePDF{l3}&\;=\;&\ePDF{\lm}+\ePDF{\lp}+\ePDF{\mu^-}+\ePDF{\mu^+}
-2\left(\ePDF{\tau^-}+\ePDF{\tau^+}\right)
& \phantom{aaa}\longleftrightarrow \phantom{aaa} &
P_{l}^+
\\
\ePDF{uc}&\;=\;&\ePDF{u}+\ePDF{\bar{u}}-\left(\ePDF{c}+\ePDF{\bar{c}}\right)
& \phantom{aaa}\longleftrightarrow \phantom{aaa} &
P_{u}^+
\\
\ePDF{ds}&\;=\;&\ePDF{d}+\ePDF{\bar{d}}-\left(\ePDF{s}+\ePDF{\bar{s}}\right)
& \phantom{aaa}\longleftrightarrow \phantom{aaa} &
P_{d}^+
\\
\ePDF{sb}&\;=\;&\ePDF{s}+\ePDF{\bar{s}}-\left(\ePDF{b}+\ePDF{\bar{b}}\right)
& \phantom{aaa}\longleftrightarrow \phantom{aaa} &
P_{d}^+
\\
\ePDF{f,{\rm\sss NS}}&\;=\;&\ePDF{f}-\ePDF{\bar{f}}
& \phantom{aaa}\longleftrightarrow \phantom{aaa} &
P_{f}^-
\end{array}
\,.
\label{tableNS}
\eeq
In general, the perturbative coefficients of the $\pm$ combinations
of the kernels stemming from eq.~(\ref{APscex}) that are relevant to
eq.~(\ref{tableNS}) are non-null, except for:
\beq
P_{l}^{\pm(1,0)}=P_{l}^{\pm(2,0)}=P_{l}^{\pm(1,1)}=0\,.
\eeq
The singlet evolution occurs in a five-dimensional space, that corresponds
to the following three linear combinations:
\beqn
\ePDF{\Sigma^l}&=&\sum_l^{\Nl}\left(\ePDF{l^-}+\ePDF{l^+}\right)\,,
\label{defSigl}
\\
\ePDF{\Sigma^u}&=&\sum_u^{\Nu}\left(\ePDF{u}+\ePDF{\bar{u}}\right)\,,
\label{defSigu}
\\
\ePDF{\Sigma^d}&=&\sum_d^{\Nd}\left(\ePDF{d}+\ePDF{\bar{d}}\right)\,,
\label{defSigd}
\eeqn
in addition to the photon ($\ePDF{\gamma}$) and gluon ($\ePDF{g}$) 
densities. It is written thus:
\beq
\frac{\partial}{\partial\log\mu^2}
\left(
\begin{array}{c}
\ePDF{\Sigma^u} \\
\ePDF{\Sigma^d} \\
\ePDF{\Sigma^l} \\
\ePDF{\gamma} \\
\ePDF{g} \\
\end{array}
\right)
=\APmat\otimes
\left(
\begin{array}{c}
\ePDF{\Sigma^u} \\
\ePDF{\Sigma^d} \\
\ePDF{\Sigma^l} \\
\ePDF{\gamma} \\
\ePDF{g} \\
\end{array}
\right),
\label{evolS}
\eeq
with the conventions of eq.~(\ref{APmatex}) for the perturbative
coefficients. Explicitly, up to the NLO we have:
\beqn
\APmat^{(1,0)}&=&\left(
\begin{array}{ccccc}
P_{q}^{+(1,0)} &
0 &
\phantom{a}0\phantom{a} &
\phantom{a}0\phantom{a} &
2\Nu P_{qg}^{(1,0)} \\
0 &
P_{q}^{+(1,0)} &
0 &
0 &
2\Nd P_{qg}^{(1,0)} \\
0 &
0 &
0 &
0 &
0 \\
0 &
0 &
0 &
0 &
0 \\
P_{gq}^{(1,0)} &
P_{gq}^{(1,0)} &
0 &
0 &
P_{gg}^{(1,0)} \\
\end{array}
\right),
\label{APmat10}
\eeqn
\beqn
\APmat^{(2,0)}&=&\left(
\begin{array}{ccccc}
P_{q}^{+(2,0)}+2\Nu P_{qq}^{{\rm S}(2,0)} &
2\Nu P_{qq}^{{\rm S}(2,0)} &
0 &
0 &
2\Nu P_{qg}^{(2,0)} \\
2\Nd P_{qq}^{{\rm S}(2,0)} &
P_{q}^{+(2,0)}+2\Nd P_{qq}^{{\rm S}(2,0)} &
\phantom{a}0\phantom{a} &
\phantom{a}0\phantom{a} &
2\Nd P_{qg}^{(2,0)} \\
0 &
0 &
0 &
0 &
0 \\
0 &
0 &
0 &
0 &
0 \\
P_{gq}^{(2,0)} &
P_{gq}^{(2,0)} &
0 &
0 &
P_{gg}^{(2,0)} \\
\end{array}
\right),
\label{APmat20}
\\
\APmat^{(0,1)}&=&\left(
\begin{array}{ccccc}
P_{u}^{+(0,1)} &
0 &
\phantom{a}0\phantom{a} &
2\Nu P_{u\gamma}^{(0,1)} &
\phantom{a}0\phantom{a} \\
0 &
P_{d}^{+(0,1)} &
0 &
2\Nd P_{d\gamma}^{(0,1)} &
0 \\
0 &
0 &
P_{l}^{+(0,1)}  &
2\Nl P_{l\gamma}^{(0,1)} &
0 \\
P_{\gamma u}^{(0,1)} &
P_{\gamma d}^{(0,1)} &
P_{\gamma l}^{(0,1)} &
P_{\gamma\gamma}^{(0,1)} &
0 \\
0 &
0 &
0 &
0 &
0 \\
\end{array}
\right),
\label{APmat01}
\\
\APmat^{(0,2)}&=&\left(
\begin{array}{ccccc}
P_{u}^{+(0,2)}+2\Nu P_{uu}^{{\rm S}(2,0)}  &
2\Nu P_{ud}^{{\rm S}(2,0)} &
2\Nu P_{ul}^{{\rm S}(2,0)} &
2\Nu P_{u\gamma}^{(0,2)} &
\phantom{a}0\phantom{a} \\
2\Nd P_{du}^{{\rm S}(2,0)} &
P_{d}^{+(0,2)}+2\Nd P_{dd}^{{\rm S}(2,0)} &
2\Nd P_{dl}^{{\rm S}(2,0)} &
2\Nd P_{d\gamma}^{(0,2)} &
0 \\
2\Nl P_{lu}^{{\rm S}(2,0)} &
2\Nl P_{ld}^{{\rm S}(2,0)} &
P_{l}^{+(0,2)}+2\Nl P_{ll}^{{\rm S}(2,0)} &
2\Nl P_{l\gamma}^{(0,2)} &
0 \\
P_{\gamma u}^{(0,2)} &
P_{\gamma d}^{(0,2)} &
P_{\gamma l}^{(0,2)} &
P_{\gamma\gamma}^{(0,2)} &
0 \\
0 &
0 &
0 &
0 &
0 \\
\end{array}
\right),
\nonumber\\*&&
\label{APmat02}
\\
\APmat^{(1,1)}&=&\left(
\begin{array}{ccccc}
P_{u}^{+(1,1)} &
0 &
\phantom{a}0\phantom{a} &
2\Nu P_{u\gamma}^{(1,1)} &
2\Nu P_{ug}^{(1,1)} \\
0 &
P_{d}^{+(1,1)} &
0 &
2\Nd P_{d\gamma}^{(1,1)} &
2\Nd P_{dg}^{(1,1)} \\
\phantom{a}0\phantom{a} &
0 &
0 &
0 &
0 \\
P_{\gamma u}^{(1,1)} &
P_{\gamma d}^{(1,1)} &
0 &
P_{\gamma\gamma}^{(1,1)} &
P_{\gamma g}^{(1,1)} \\
P_{gu}^{(1,1)} &
P_{gd}^{(1,1)} &
0 &
P_{g\gamma}^{(1,1)} &
P_{gg}^{(1,1)} \\
\end{array}
\right).
\label{APmat11}
\eeqn
The entries of the matrices in eqs.~(\ref{APmat10})--(\ref{APmat11})
can be found in, or easily derived from the results of, 
refs.~\cite{Curci:1980uw,Furmanski:1980cm,Ellis:1996nn,deFlorian:2015ujt,
deFlorian:2016gvk}.

As is shown in refs.~\cite{Bertone:2019hks,Frixione:2021wzh,Bertone:2022ktl},
the evolution equation of eq.~(\ref{evolS}) (and analogously for its 
non-singlet counterparts of eq.~(\ref{evolNS})) is conveniently solved 
in Mellin space, where it is re-expressed in terms of an evolution
operator, thus:
\beq
\frac{\partial \Eop_{N}^{(K)}(\mu)}{\partial\log\mu^2} =
\Mmat_{N}^{(K)}(\aem(\mu),\as(\mu))\,\Eop_{N}^{(K)}(\mu)\,,
\label{evEopS}
\eeq
where the index $K$ understands a dependence on the chosen factorisation
scheme (see refs.~\cite{Frixione:2019lga,Frixione:2021wzh} for more
details). We stress that the evolution kernel $\Mmat_{N}^{(K)}$
coincides with the Mellin transform of the matrix $\APmat$ that
appears in eq.~(\ref{evolS}) (or with its scalar counterparts
$P_a$ that appear in eq.~(\ref{evolNS}) for the non-singlet densities)
{\em only} if the factorisation and renormalisation schemes are both
chosen equal to $\MSb$ -- in all of the other cases, $\Mmat_{N}^{(K)}$
is constructed starting from $\APmat$ following the procedure 
described in refs.~\cite{Frixione:2021wzh,Bertone:2022ktl}.
Furthermore, the kernel $\Mmat_{N}^{(K)}$ depends on parameters whose
values are determined by the mass range one considers, and specifically
by the number of lepton and light-quark flavours which are active there.
Therefore, eq.~(\ref{evEopS}) is solved in the range \mbox{$\mu_I<\mu<m_k$},
with $m_k$ a fermion threshold, and \mbox{$\mu_I=\max(\mu_0,m_{k-1})$},
starting with a $k$ value for which $\mu_I=\mu_0$, and with continuity 
conditions imposed at the thresholds.

As far as such a solution is concerned, we note that eq.~(\ref{evEopS}) 
has the same structure as its counterpart in ref.~\cite{Bertone:2022ktl}, 
except for the fact that it differs from the latter in a single 
practical aspect. Namely, in ref.~\cite{Bertone:2022ktl}
we have considered the cases where $\Mmat_{N}^{(K)}$ does not depend on 
$\mu$, neither directly nor through $\aem(\mu)$ (this corresponds to the 
$\MSb$ UV-renormalisation scheme where the running of $\aem$ is neglected);
where $\Mmat_{N}^{(K)}$ depends on $\mu$ only through $\aem(\mu)$ 
(this corresponds to the $\MSb$ UV-renormalisation scheme); and finally where
$\Mmat_{N}^{(K)}$ has a direct dependence on $\mu$, but $\aem$ does not run
(this corresponds to the $\aem(\mZ)$ and $G_\mu$ UV-renormalisation schemes).
Thus, here we have to deal with the last possible case not already taken 
into account in ref.~\cite{Bertone:2022ktl}, namely that where 
$\Mmat_{N}^{(K)}$ has a direct dependence on $\mu$ (through $\as(\mu)$), 
and $\aem$ is running. It turns out that we can still solve the corresponding 
evolution equation in a similar manner as in ref.~\cite{Bertone:2022ktl}, 
i.e.~by relying on a path-ordered approach. However, we do so in two different
ways, that correspond to choosing different evolution variables; the 
relevant solutions are then used to validate each other by means 
of mutual cross-checks. The first evolution variable we adopt is:
\beq
l = \log\frac{\mu^2}{\mu^2_I}\,;
\label{ldef}
\eeq
as was already said, $\mu_I$ is the lower end of the current mass range.  
For each value of $l$, we obtain $\mu$ by simply inverting eq.~(\ref{ldef}), 
i.e.~\mbox{$\mu^2 = \mu_I^2\,\exp(l)$}, and we then evaluate both $\aem(\mu)$ 
and $\as(\mu)$. The second evolution variable is:
\beq
t=\frac{1}{2\pi \bzQED}\log\frac{\aem(\mu)}{\aem(\mu_I)}\,,
\label{tdef}
\eeq
with $\bzQED$ the LO coefficient of the QED $\beta$ function, in the range
where the evolution is taking place, defined according to what follows:
\beq
\frac{d\aem(\mu)}{d\log\mu^2}\equiv\beta_{\QED}(\aem) = 
\bzQED \aem^2 + \boQED \aem^3 + \ord(\aem^4)\,.
\label{eq:QEDRGE}
\eeq
In this case, we can immediately obtain $\aem(\mu)$ by inverting
eq.~(\ref{tdef}):
\beq
\aem(\mu) = \aem(\mu_I)\,e^{2\pi \bzQED\,t}\,.
\eeq
In order to obtain $\as(\mu)$ for any $\mu>\mu_I$, we first solve the 
QED RGE eq.~(\ref{eq:QEDRGE}) to obtain $\mu$ from $\aem(\mu)$
\beq
\mu^2 = \mu_I^2\exp\left[\!B_{\QED}\Big(\aem(\mu)\Big) - 
B_{\QED}\Big(\aem(\mu_I)\Big)\right],
\eeq
with the $B_{\QED}$ function at the NLO given by
\beq
B_{\QED}(a)\equiv
\int \frac{da}{\beta_{\QED}(a)} =
- \frac{1}{\bzQED a} - \frac{\boQED}{\left(\bzQED\right)^2}
\log\left(\frac{a}{\bzQED + a\,\boQED}\right)\,,
\eeq
and we then evaluate $\as(\mu)$.

The path-ordered approach relies on a discretised solution, which is
explained in details in sect.~5.2. of ref.~\cite{Bertone:2022ktl}. Here, 
we limit ourselves to mentioning that, in view of the more complicated 
scale dependence of $\Mmat_{N}^{(K)}$ w.r.t.~that of the cases studied
in ref.~\cite{Bertone:2022ktl}, and of the more extended scale range
w.r.t.~that relevant to the electron PDFs, we have re-assessed the 
dependence of the PDFs upon the number of discrete steps by means of 
which the evolution operator is constructed. Denoting by $n_k$ the
number of discrete steps in the $k^{th}$ mass-threshold interval,
and by \mbox{$n=\sum_k n_k$} the total number of steps, we have found
very stable solutions with \mbox{$50\le n_k\le 1000$} and
\mbox{$1000\le n\le 2000$}: with these ranges, the muon and photon
PDFs exhibit a relative variation of $\ord(10^{-7})$, while for all
of the other partons the effect is of $\ord(10^{-5})$. As such, the
impact of the variations of these discrete parameters on the PDFs is
totally negligible; we point out that this also allows us to conclude that
the inclusion of the scale dependence of $\as$ does not lead to any
qualitative or quantitative changes w.r.t.~the findings of
ref.~\cite{Bertone:2022ktl}.

As was already mentioned, for the phenomenological results presented
in this paper we retain only the $\ord(\as)$ and $\ord(\aem)$ kernels,
i.e.~those of eqs.~(\ref{APmat10}) and~(\ref{APmat01}) respectively.
We also point out that the basis of the non-singlet functions in
eq.~(\ref{tableNS}) is only one among several equally sensible choices;
we have verified, by also adopting the basis suggested in
ref.~\cite{Bertone:2015lqa}, that our results are basis-independent.

\section{Results\label{sec:res}}
In this section we present our predictions for the muon PDFs, and
for the dijet cross sections that result from them. In particular,
we are interested in assessing the uncertainties that stem from 
the parametrisation of the strong coupling constant at small
scales. In order to do this, in addition to our own approach
which has been discussed in detail in sect.~\ref{sec:QCD}, we
have also implemented the strategy of refs.~\cite{Han:2020uid,
Han:2021kes,Garosi:2023bvq}, where $\as$ is set equal to zero for
\mbox{$\mu\le Q_0$}. We shall refer to these two approaches
as ``analytical'' and ``truncated'', respectively. In both of these
cases, the uncertainties are defined as the envelopes of the
variations of the relevant low-energy parameter(s) -- for the analytical
approach, these are given in table~\ref{tab:bcppars}, whereas for
the truncated approach we take 
\mbox{$Q_0\in [0.52,1.0]$~GeV}~\cite{Han:2021kes,Garosi:2023bvq}.
We point out that there is a fundamental difference between the
uncertainties thus computed with the two methods. Namely, while
the low-energy-parameter ranges are to a certain extent always 
arbitrary, the lower end of the $Q_0$ range is necessarily an unstable
point, since small variations towards lower values induce progressively
larger envelopes (as one moves towards the Landau pole of the 
perturbatively-computed $\as$); this behaviour has no analogue
in the analytical approach we employ in this work.

\subsection{Results for the PDFs\label{sec:resPDF}}
In this section we focus on the PDFs as standalone objects.
In general, the muon PDF is utterly dominant, and its size dwarfs that 
of the PDFs of all of the other partons; in this sense, the situation is 
quite analogous to that of the electron PDFs. However, this statement
is only correct at intermediate and large $z$ values: at $z\to 0$,
other partons dominate. As was already discussed in sect.~\ref{sec:intro}, 
the large c.m.~energies expected at muon colliders render the $z\to 0$
region quite relevant for certain production processes. Therefore,
in view of the fact that this constitutes a novelty w.r.t.~what
happens at $\epem$ colliders, we shall concentrate on this $z$ region
in the following. However, we stress that our PDFs do treat the large-$z$
region without any approximation or discontinuity, and as their
electron counterparts~\cite{Bertone:2019hks,Bertone:2022ktl} they can
be straightforwardly used to study the production of large-mass systems
as well.

Before proceeding to show explicit results, we mention a couple of
features of the PDFs (with the analytical low-scale approach) that
serve as cross checks. Firstly, we find that by not including the
$\ord(\as)$ splitting kernels (i.e.~by performing a pure-QED LL evolution,
with all of the partons except for the gluon, which cannot be generated
in this way) the PDFs of all the leptons and of the photon change by a 
negligible amount (a relative $\ord(10^{-5})$ factor or smaller) w.r.t.~to our 
default $\ord(\as)+\ord(\aem)$ coupled-evolution PDFs. Conversely, the PDFs 
of the quarks display relative differences of tens of percent; typically,
the inclusion of QCD splitting kernels results in enhancing (depleting)
the PDFs at small (large) $z$ values. Overall, and consistently with
basic expectations, the enhancement has a much larger effect than the
depletion, since quark PDFs decrease with $z$ (see later).
Secondly, while by default we use two-loop coupling constants in our 
evolution, we have verified that, by employing a one-loop expression for 
$\as$, the lepton and photon PDFs change by a relative $\ord(10^{-6})$ 
factor. While the effect on the quark and gluon PDFs is much larger
(up to 10\% at $\mu=30$~GeV), it is basically driven by the differences
between the two RGE solutions for $\as$ (at $\mu=m_c$, the two-loop value
of $\as(\mu)$ is about 17\% larger than its one-loop counterpart, if
an initial condition $\as(\mZ)=0.118$ is adopted for both evolutions).

\begin{table}[th!]
  \begin{center}    
    \begin{tabular}{|c|rr|}
      \hline
  & an.$\phantom{aaaa}$ & tr.$\phantom{aaaa}$ \\
      \hline
$\mu^-$  & $97.8592_{-0.00\%}^{+0.00\%}$ & $97.8592_{-0.00\%}^{+0.00\%}$ \\
$\gamma$ &  $2.0756_{-0.00\%}^{+0.00\%}$ &  $2.0756_{-0.00\%}^{+0.00\%}$ \\
$\sum l$ &  $0.0265_{-0.00\%}^{+0.00\%}$ &  $0.0265_{-0.00\%}^{+0.00\%}$ \\
$\sum q$ &  $0.0332_{-0.02\%}^{+0.02\%}$ &  $0.0333_{-0.23\%}^{+0.17\%}$ \\
$g$      &  $0.0056_{-0.10\%}^{+0.10\%}$ &  $0.0055_{-1.03\%}^{+1.01\%}$ \\
     \hline
    \end{tabular}
  \end{center}
  \caption{
\label{tab:momLL} Fractions of momentum carried by the muon, the photon,
all of the leptons different from $\mu^-$ ($\sum l$), all of the quarks
($\sum q$), and the gluon, computed with the PDFs based on the analytical
(left column) and truncated (right column) low-energy approach. The
results are obtained at $\mu=m_{\sss Z}$.
}
\end{table}
In table~\ref{tab:momLL} we report the fractions of muon momentum carried
by either individual partons or combinations of partons, at $\mu=m_{\sss Z}$.
In addition to the central values, we show the fractional uncertainties
due to the choices of the low-energy parameters. We note that the 
central values stemming from the analytical and truncated approaches
are essentially identical or very close to each other. In fact, this
is an artifact of the definition of the momentum fraction, that
weights each PDF by a factor of $z$, which thus suppresses the contributions
of the small-$z$ region; as we shall show later, the differences between
the two approaches increase with decreasing $z$. One can also observe
a reasonable agreement with the results in the last line of table~1
of ref.~\cite{Han:2021kes}, where the same quantities are reported.
As far as the uncertainties associated with the low-energy parameters
are concerned, for both the analytical and the truncated approach they
are negligible for the muon, photon, and lepton contributions. They
remain relatively small in the case of the quarks and the gluon; however, 
what one can see there is that the uncertainties of the truncated approach
are about a factor of ten larger than those stemming from the analytical
approach, which is an example of the general features discussed at
the beginning of sect.~\ref{sec:res}.

\begin{figure}[thb]
  \begin{center}
    \includegraphics[width=0.49\textwidth]{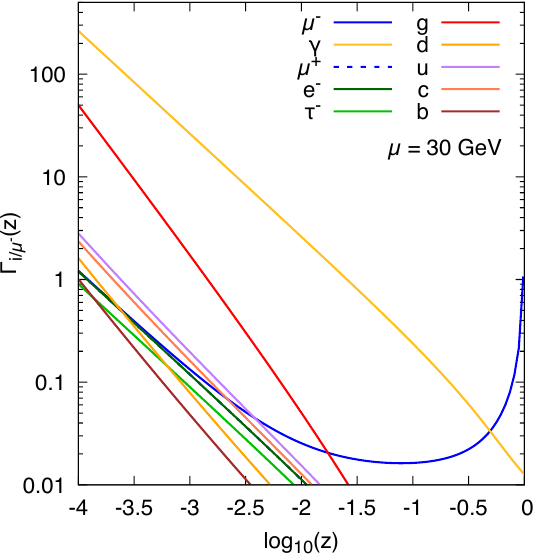}\,
    \includegraphics[width=0.49\textwidth]{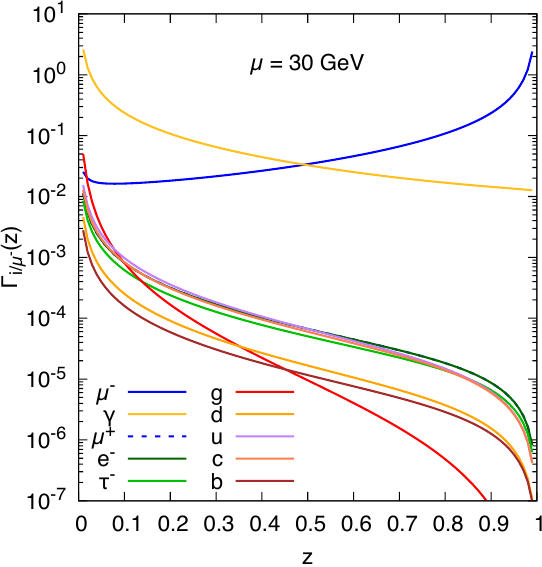}
    \caption{
\label{fig:PDF30} PDFs at $\mu=30$~GeV. The contents of the two panels are
identical, the only difference between the two being the variable on the
$x$ axis. Apart from the case of $\mu^+$, the antifermion PDFs coincide
with those of the corresponding fermions, and are not shown. Also not shown
is the strange PDF, since it coincides with that of the down. Finally, on these
scales the $\mu^+$ and $e^-$ results cannot be distinguished from one another.}
  \end{center}
\end{figure}
\begin{figure}[thb]
  \begin{center}
    \includegraphics[width=0.49\textwidth]{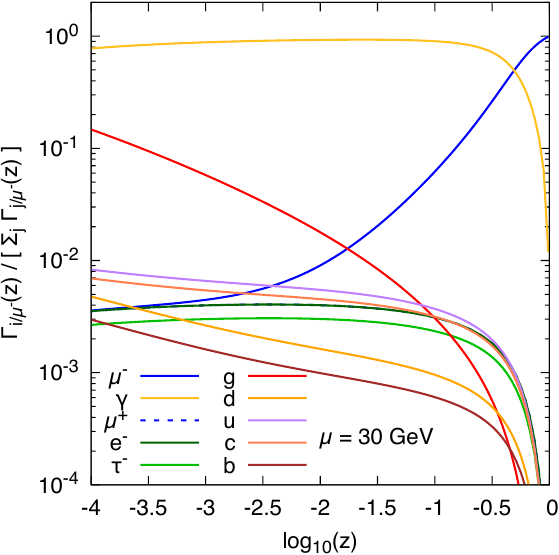}\,
    \includegraphics[width=0.49\textwidth]{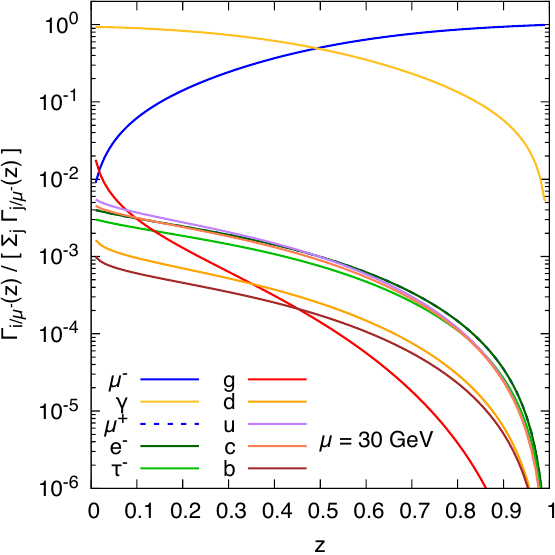}
    \caption{
\label{fig:rPDF30} Ratios of individual PDFs over their sum, at $\mu=30$~GeV. 
The contents of the two panels are identical, the only difference between 
the two being the variable on the $x$ axis. Apart from the case of $\mu^+$, 
the antifermion PDFs coincide with those of the corresponding fermions, and 
are not shown. Also not shown is the strange PDF, since it coincides with 
that of the down. Finally, on these scales the $\mu^+$ and $e^-$ results 
cannot be distinguished from one another.}
  \end{center}
\end{figure}
In fig.~\ref{fig:PDF30} we show the PDFs of all partons at $\mu=30$~GeV
(as a representative scale relevant to the production of a small-mass
system) as a function of either $\log_{10}z$ (left panel) or $z$ (right
panel). The PDFs are obtained with the analytical low-energy approach,
and correspond to the default low-energy parameters. The relative impact
of these PDFs is presented in fig.~\ref{fig:rPDF30}, where we plot the
ratios of the individual PDFs over the sum of all of them. The plots
show clearly the dominance of the muon PDF as $z\to 1$. Conversely, as
one moves towards $z=0$, all of the other partons become increasingly
important (bar for the non-muon leptons) -- still, the muon PDF remains 
larger than the photon PDF for $z\gtrsim 0.5$, larger than the gluon PDF 
for $z\gtrsim 0.017$, and larger than the PDF of the largest among the quarks 
(the up quark) for $z\gtrsim 0.004$. At small $z$ the largest PDF is that 
of the photon; however, the gluon has the steepest slope of all partons, 
and in particular its PDF increases faster than the photon one for $z\to 0$.
The cumulative impact of the quark PDFs is also non negligible in that
region. These facts, coupled with the trivial observation that short-distance 
cross sections mediated by strong interactions are much larger that those 
stemming from electroweak interactions, imply that a muon collider
constitutes a very favourable environment for clean studies of 
low-mass hadron-induced systems.

We have considered the same plots as in figs.~\ref{fig:PDF30} 
and~\ref{fig:rPDF30} for several scales, up to 10~TeV. Qualitatively, we have 
found the same patterns as at $30$~GeV; quantitatively, in the small-$z$ 
region the relative impact of the PDFs of the strongly-interacting partons 
grows with the scale. Having said that, we point out that, given the collider 
energy, larger scales are mostly associated with larger $z$ values; when such 
scales are relevant, in order to probe small $z$'s one needs to focus on
very boosted systems.

\begin{figure}[thb]
  \begin{center}
    \includegraphics[width=0.49\textwidth]{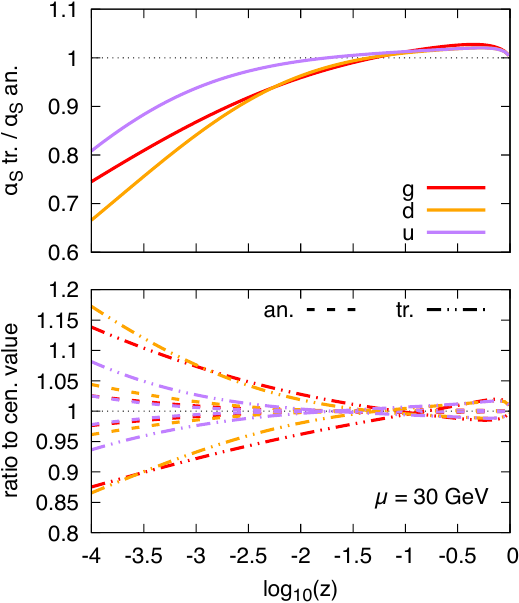}\,
    \includegraphics[width=0.49\textwidth]{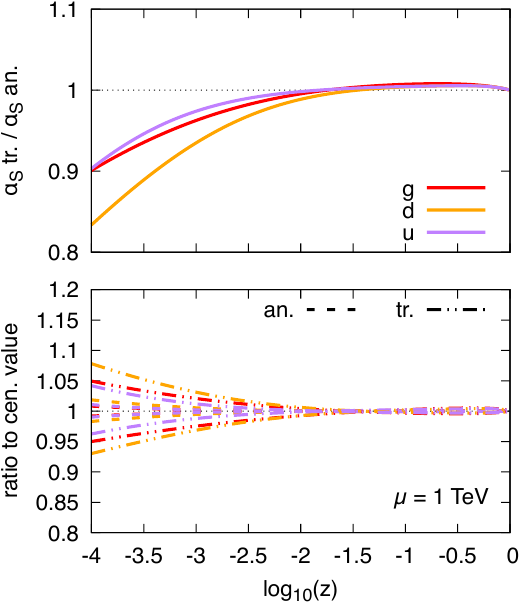}
    \caption{
\label{fig:PDFunc} Dependence of the gluon, $u$-quark, and $d$-quark PDFs
on the low-energy parameters. See the text for details.
}
  \end{center}
\end{figure}
We now turn to consider the dependence of the PDFs on the approach chosen
for the small-scale behaviour of $\as$, as well as that on the relevant
low-energy parameters. We summarise our findings in fig.~\ref{fig:PDFunc};
the left-side panels, obtained at $\mu=30$~GeV, have the same layouts as 
the right-side ones, obtained at $\mu=1$~TeV. The upper panels show the
ratio of the PDFs obtained with the default parameters in the truncated
approach over their counterparts in the analytical approach. While there
are differences among the partons considered, broadly speaking the analytical
approach gives larger (smaller) PDFs than the truncated one for
$z\lesssim 0.05$ ($z\gtrsim 0.05$). These differences are largest
at the smallest $z$ values: e.g.~for \mbox{$z=10^{-4}$}, the 
analytical-approach PDFs are larger than the truncated-approach ones
by \mbox{$20-30$\%} at $\mu=30$~GeV, and by \mbox{$10-15$\%} at $\mu=1$~TeV.
This information has to be taken together with that on the uncertainties 
due to the choices of the low-energy parameters. In the lower panels 
fig.~\ref{fig:PDFunc} we display such fractional uncertainties for
the analytical (dashed curves) and truncated (dot-dashed curves)
approaches; the curves represent the envelopes, computed as was
explained at the beginning of sect.~\ref{sec:res}. One sees here the 
local (in $z$) version of what has been observed in table~\ref{tab:momLL}:
the analytical approach gives significantly smaller uncertainties than
the truncated one. In both cases, the uncertainties decrease with
the scale. We shall soon see (sect.~\ref{sec:res2j}) the implications
of these facts on selected observables.

\subsection{Results for dijet cross sections\label{sec:res2j}}
In this section we report our predictions for dijet total rates.
We work with LO short-distance cross sections (i.e.~at
\mbox{$\ord(\as^n\aem^m)$} with \mbox{$n+m=2$}); as such, jets
coincide with partons (i.e.~there is no need for a jet-finding
algorithm), and are defined by means of the simplest acceptance cuts:
\beq
\pt^{(j)}>\pt^{cut}\,,\;\;\;\;\;\;\;\;
\abs{\eta^{(j)}}<3\,.
\label{jetcuts}
\eeq
The collider energy is $\sqrt{s}=10$~TeV, and we consider
\mbox{$\pt^{cut}=10$~GeV} and \mbox{$\pt^{cut}=100$~GeV};
the factorisation and renormalisation scales are set equal
to $\pt^{(j)}$.

We point out that in the presence of EW interactions the definition
of the partonic content of a jet constitutes a non-trivial problem
(see e.g.~ref.~\cite{Frederix:2016ost} for an extended discussion
of this matter). By working at the LO most of the complications can 
be avoided, and we can obtain sensible results by considering only 
light quarks and gluons when defining the jets. Conversely, since we 
are working at an $\mpmm$ collider, the initial-state and intermediate 
partons can be light quarks, gluons, leptons, and photons. In summary,
the results of this section have been obtained by employing the
$2\to 2$ matrix elements relevant to the processes:
\beq
p_1+p_2\;\longrightarrow\;p_3+p_4\,,
\eeq
with
\beq
p_1\,,p_2\in\big\{q_i,\bq_i,g,l_i^\pm,\gamma\big\}\,,\;\;\;\;\;\;\;\;
p_3\,,p_4\in\big\{q_i,\bq_i,g\big\}\,.
\eeq
In view of this, the intermediate $s$- or $t$-channel partons are
light quarks, gluons, or photons.

\begin{table}[th!]
 \begin{center}    
   \begin{tabular}{|c|rr|}
     \hline
$\sigma$($\pt^{cut}=10$~GeV) [pb] & an.$\phantom{aaaa}$ &
tr.$\phantom{aaaa}$ \\
     \hline
$\ord(\as^2)$ & $18.33_{-1.25\%}^{+1.30\%}$ & 
                $15.00_{-10.99\%}^{+10.23\%}$ \\
$\gamma$-ind. & $8.24_{-0.91\%}^{+0.68\%}$ &
                $7.56_{-3.75\%}^{+3.71\%}$ \\
total         & $26.58_{-1.15\%}^{+1.11\%}$ &
                $22.57_{-8.56\%}^{+8.04\%}$ \\
     \hline
    \end{tabular}
  \end{center}
 \caption{
\label{tab:dijets10} Total dijet rates for $\pt^{cut}=10$~GeV,
in pb.
}
\end{table}
\begin{table}[th!]
 \begin{center}    
   \begin{tabular}{|c|rr|}
     \hline
$\sigma$($\pt^{cut}=100$~GeV) [fb] & an.$\phantom{aaaa}$ &
tr.$\phantom{aaaa}$ \\
     \hline
$\ord(\as^2)$ & $41.38_{-0.03\%}^{+0.03\%}$ &
                $41.17_{-0.85\%}^{+0.46\%}$ \\
$\gamma$-ind. & $90.03_{-0.02\%}^{+0.01\%}$ &
                $89.67_{-0.32\%}^{+0.24\%}$ \\
total         & $136.91_{-0.00\%}^{+0.00\%}$ &
                $136.35_{-0.48\%}^{+0.28\%}$ \\
     \hline
    \end{tabular}
  \end{center}
 \caption{
\label{tab:dijets100} Total dijet rates for $\pt^{cut}=100$~GeV,
in fb.
}
\end{table}
The total dijet rates within the acceptance cuts of eq.~(\ref{jetcuts}) are 
given in the last line of tables~\ref{tab:dijets10} (for $\pt^{cut}=10$~GeV) 
and~\ref{tab:dijets100} (for $\pt^{cut}=100$~GeV) -- note the different
units employed in these two tables. In addition to these, we also report
the results for the sum of the $\ord(\as^2)$ contributions (first line), 
i.e.~of the pure-QCD processes, and for the sum of the photon-induced 
contributions (second line), i.e.~those with $\gamma\gamma$, $\gamma q$, 
and $\gamma g$ partonic initial states. Thus, the differences between
the results in the third line and the sum of their counterparts in the first
and second lines is equal to the sum of the $\ord(\aem^2)$ contributions 
of the four-fermion processes.

These results complement, and confirm, those for standalone PDFs that
we have discussed in sect.~\ref{sec:resPDF}. In particular: 
{\em a)} the low-energy-parameter uncertainties of the analytical
approach are always much smaller than those of the truncated one;
{\em b)} these uncertainties are significantly larger with
$\pt^{cut}=10$~GeV than with $\pt^{cut}=100$~GeV; in the latter case,
those stemming from the analytical approach are essentially
negligible\footnote{Note that there the uncertainty on the total is
smaller than that on its major components, i.e.~the $\ord(\as^2)$ and
photon-induced contributions. This is due to the fact that, with such
small numbers, the usual envelope computation with discrete sets of
parameters would have to be promoted to a continuous parameter scan.};
{\em c)} at $\pt^{cut}=10$~GeV, the central values stemming from the
analytical and truncated approaches are fairly different from one another. 
Moreover, they are not within the low-energy parameter uncertainty
ranges; this underscores again the fundamental difference between the
two approaches, which plays an increasingly important role as one
probes smaller $z$ values, i.e.~not only for small-mass systems, but
also for those that feature large longitudinal boosts. 
We conclude with two additional observations. Firstly, the low-energy 
parameter uncertainties of the photon-induced contributions stem entirely 
from the corresponding quark- and gluon-PDF uncertainties (i.e.~these are
negligible in the case of the $\gamma\gamma$ channel; as it was discussed
sect.~\ref{sec:resPDF}, the photon PDF is fairly stable against the
variations of these parameters). Secondly, the relative impact of the
four-fermion contributions grows with $\pt^{cut}$: it is negligible at
$\pt^{cut}=10$~GeV, and becomes equal to about 4\% of the total at
$\pt^{cut}=100$~GeV, where it is overwhelmingly due to the $l^+l^-$
channels.

\section{Conclusions and outlook\label{sec:concl}}
We have computed the PDFs of the unpolarised muon in the context of 
a coupled QED-QCD evolution; the set of partons is constituted
by the leptons, the light quarks (including the $c$ and $b$ quarks),
the photon, and the gluon. Our work differs in several notable ways 
w.r.t.~those in the literature~\cite{Han:2020uid,Han:2021kes,Garosi:2023bvq} 
that have studied the same topic. Firstly, we adopt a dispersive-inspired
parametrisation of $\as$ that allows one to compute it for any scale
values, including those of $\ord(\LQCD)$ and below. This implies that
the resulting predictions are stable w.r.t.~the variations of the
low-energy parameters that control the small-scale behaviour of $\as$,
at variance with what happens in the truncated approach of
refs.~\cite{Han:2020uid,Han:2021kes,Garosi:2023bvq}, which is
inherently unstable. An additional benefit of our approach is that 
its low-energy parameters can be fitted to data (typically, of
$\epem$ event shapes or of hadronic structure functions) without loss
of predictive power in the context of $\mpmm$ physics, since they
are assumed to underpin the universal features of small-scale $\as$.
Secondly, we do not consider the heavy electroweak vector bosons as
partons. In essence, even for collider energies of tens of TeVs,
the cases where the boson masses are negligible, and the resummation
of the large logarithms they induce is important, are extremely marginal; 
conversely, vector-boson mass effects are ubiquitous, and this suggests
that a treatment based on perturbative matrix-element computations is
to be preferred. Thirdly, while in this paper we limit ourselves to
presenting LL-accurate predictions, in order to facilitate the comparison
with the results of refs.~\cite{Han:2020uid,Han:2021kes,Garosi:2023bvq}
in terms of the aspects mentioned above, we have shown that an increase
of precision does not entail any new conceptual features, and can thus
be easily achieved.

From a practical viewpoint, the muon PDFs are computed in the 
same manner as was done for their electron counterparts in 
refs.~\cite{Bertone:2019hks,Bertone:2022ktl}: in particular, they take 
into account all lepton and light-quark mass thresholds, and are continuous
and smooth over the whole \mbox{$z\in (0,1)$} range. As such, they do
not present any qualitatively new feature in the $z\to 1$ region,
where the muon-parton PDF is hugely dominant. Because of this, in our
phenomenological studies we have only considered the small-$z$ region,
either by looking directly at the PDFs, or by computing dijet cross
sections with small transverse-momentum cuts. The overall conclusion is 
that the different treatment of the small-scale behaviour of $\as$
w.r.t.~that of refs.~\cite{Han:2020uid,Han:2021kes,Garosi:2023bvq}
induces visible difference both for the central values and for the
uncertainties associated with low-energy parameter variations, which
in our case are significantly smaller.

As a follow-up of this work we shall compute the PDFs at the NLL
accuracy in the QED, QCD, and mixed QED$\times$QCD contributions.
While interesting in itself, the crucial advantage of our ability
to do so is that it gives us the possibility of studying in more 
detail the small-$z$ behaviour of the PDFs and, if need be, to
resum the large logarithmic contributions that potentially arise there.

\section*{Acknowledgements}
We are grateful to V.~Bertone, M.~Bonvini, M.~Dasgupta, F.~Maltoni,
B.~Webber, and M.~Zaro for several useful conversations. SF thanks the 
CERN TH division for the kind hospitality during the course of this work.
GS is supported in part by the Swiss National Science Foundation (SNF)
under contract 200020-204200.

\phantomsection
\addcontentsline{toc}{section}{References}
\bibliographystyle{JHEP}
\bibliography{mupdfs}

\end{document}